\documentclass[a4paper,twocolumn,11pt,accepted=2024-12-19]{quantumarticle}
\pdfoutput=1
\usepackage[utf8]{inputenc}
\usepackage[english]{babel}
\usepackage[T1]{fontenc}
\usepackage{amsmath}
\usepackage{hyperref}

\usepackage{tikz}
\usepackage{lipsum}

\usepackage{bm}
\usepackage{multirow}

\usepackage{siunitx}
\DeclareSIUnit{\sample}{Sa}
\DeclareSIUnit{\baud}{Baud}
\DeclareSIUnit{\snu}{SNU}
\DeclareSIUnit{\frame}{frame}

\usepackage{glossaries-extra}
\setabbreviationstyle[acronym]{long-short}
\newacronym{qkd}{QKD}{Quantum Key Distribution}
\newacronym{cvqkd}{CV-QKD}{Continuous-Variable Quantum Key Distribution}
\newacronym{dvqkd}{DV-QKD}{Discrete-Variable Quantum Key Distribution}
\newacronym{mbc}{MBC}{Modulator Bias Controller}
\newacronym{voa}{VOA}{Variable Optical Attenuator}
\newacronym{dac}{DAC}{Digital-to-Analog Converter}
\newacronym{dsp}{DSP}{Digital Signal Processing}
\newacronym{lo}{LO}{Local Oscillator}
\newacronym{llo}{LLO}{Local Local Oscillator}
\newacronym{bhd}{BHD}{Balanced Homodyne Detector}
\newacronym{adc}{ADC}{Analog-to-Digital Converter}
\newacronym{pm}{PM}{Polarisation Maintaining}
\newacronym{psk}{PSK}{Phase-Shift Keying}
\newacronym{qam}{QAM}{Quadrature Amplitude Modulation}
\newacronym{pcs}{PCS}{Probabilistic Constellation Shaping}
\newacronym{gmcs}{GMCS}{Gaussian Modulated Coherent States}
\newacronym{cazac}{CAZAC}{Constant Amplitude Zero AutoCorrelation}
\newacronym{rc}{RC}{Raised Cosine}
\newacronym{rrc}{RRC}{Root-Raised Cosine}
\newacronym{ossb}{OSSB}{Optical Single SideBand}
\newacronym{skr}{SKR}{Secret Key Rate}
\newacronym{ldpc}{LDPC}{Low Density Parity Check}
\newacronym{hal}{HAL}{Hardware Abstraction Layer}
\newacronym{gui}{GUI}{Graphical User Interface}
\newacronym{snr}{SNR}{Signal-to-Noise ratio}
\newacronym{isi}{ISI}{Inter-Symbol Interference}
\newacronym{qpsk}{QPSK}{Quadrature Phase Shift Keying}
\newacronym{snu}{SNU}{Shot Noise Units}
\newacronym{pbs}{PBS}{Polarisation Beam Splitter}
\newacronym{qrng}{QRNG}{Quantum Random Number Generator}

\begin{document}

\title{QOSST: A Highly-Modular Open Source Platform for Experimental Continuous-Variable Quantum Key Distribution}

\author{Yoann Piétri}
\affiliation{Sorbonne Université, CNRS, LIP6, F-75005 Paris, France}
\orcid{0009-0005-0734-3529}
\email{Yoann.Pietri@lip6.fr}
\author{Matteo Schiavon}
\orcid{0000-0002-4631-457X}
\affiliation{Sorbonne Université, CNRS, LIP6, F-75005 Paris, France}
\author{Valentina Marulanda Acosta}
\affiliation{Sorbonne Université, CNRS, LIP6, F-75005 Paris, France}
\affiliation{DOTA, ONERA, Université Paris Saclay, F-92322 Châtillon, France}
\orcid{‪0000-0002-2131-2172‬}

\author{Baptiste Gouraud}
\affiliation{Exail, F-25000, Besançon}
\orcid{0000-0002-5640-9150}

\author{Luis Trigo Vidarte}
\affiliation{ICFO - Institut de Ciènces Fotòniques, The Barcelona Institute of Science and Technology, Castelldefels (Barcelona) 08860, Spain}
\orcid{0000-0003-3686-3820}

\author{Philippe Grangier}
\affiliation{Université Paris-Saclay, Institut d'Optique Graduate School, CNRS, Laboratoire Charles Fabry, 91127, Palaiseau, France}

\author{Amine Rhouni}
\affiliation{Sorbonne Université, CNRS, LIP6, F-75005 Paris, France}
\orcid{0000-0002-2865-3674}

\author{Eleni Diamanti}
\affiliation{Sorbonne Université, CNRS, LIP6, F-75005 Paris, France}
\orcid{0000-0003-1795-5711}
\maketitle

\begin{abstract}
Quantum Key Distribution (QKD) enables secret key exchange between two remote parties with information-theoretic security rooted in the laws of quantum physics. Encoding key information in continuous variables (CV), such as the values of quadrature components of coherent states of light, brings implementations much closer to standard optical communication systems, but this comes at the price of significant complexity in the digital signal processing techniques required for operation at low signal-to-noise ratios. In this work, we wish to lower the barriers to entry for CV-QKD experiments associated to this difficulty by providing a highly modular, open source software that is in principle hardware agnostic and can be used in multiple configurations. We benchmarked this software, called QOSST, using an experimental setup with a locally generated local oscillator, frequency multiplexed pilots and RF-heterodyne detection, and obtained state-of-the-art secret key rates of the order of Mbit/s over metropolitan distances at the asymptotic limit. We hope that QOSST can be used to stimulate further experimental advances in CV-QKD and be improved and extended by the community to achieve high performance in a wide variety of configurations.   
\end{abstract}

\section{Introduction}

\gls{qkd} aims at the exchange of a random string of bits --~a cryptographic key~-- between two trusted users, commonly referred to as Alice and Bob, who are linked through an untrusted quantum channel and a public authenticated classical channel. This random bit string can then be used in symmetric encryption ciphers. Crucially, the security of the key exchange is not based on computational assumptions but on the laws of quantum physics leading to information-theoretic security~\cite{AdvancesInQuaPirand2020}.


Two major families of QKD protocols can be distinguished by the degrees of freedom in which the key information is encoded. In \gls{dvqkd}, these are properties of single photons, such as polarisation or phase~\cite{BennettBrassard84,OvercomingTheLucama2018}, while in \gls{cvqkd}, they are continuous degrees of freedom, such as the quadratures of coherent states~\cite{ContinuousVariGrossh2002}. In terms of practical implementation, an important difference between these two families lies in the detection technique. \gls{dvqkd} systems require single-photon detectors, which have limited detection rates due to their dead time and need to be operated in sub-kelvin temperatures to achieve a high efficiency~\cite{Ip:08, InvitedReviewEisama2011}. \gls{cvqkd} systems, on the other hand, use coherent optical detectors at room temperature. This reduces the size and cost of the detection apparatus and enables higher bandwidth detectors and hence higher measurement rates. Such detectors are also easier to integrate~\cite{AdvancesInChiLiuQ2022}.

Several experimental demonstrations of \gls{cvqkd} have been performed in the past few years, including over long distances~\cite{LongDistanceCZhang2020}, with high key rates~\cite{ExperimentalDePanY2022,ExperimentalDeRoumes2024} and with integrated transmitters or receivers~\cite{Aldama:23,Pietri:24,Hajomer:24}. For such protocols that are inspired by classical communication systems, an important part of the implementation complexity comes down to the \gls{dsp} techniques required to recover the symbols at Bob's side and to correct the impairments coming from the transmission at a low \gls{snr}. Indeed, to limit the amount of information leaked to the eavesdropper, Eve, \gls{cvqkd} systems operate in the regime of a few photons per symbol, which is the key difference with classical coherent systems that operate at high \gls{snr}. This is common to the several variations of \gls{cvqkd} protocols that exist, in terms of symbol shaping, generating a \gls{lo} for the coherent detection, or sending classical information for frequency and phase compensation. 

In this work, our main contribution is to provide an open source, highly modular and highly customizable platform for \gls{cvqkd}, that we call QOSST: Quantum Open Software for Secure Transmissions. In practice, our platform is capable of performing \gls{cvqkd} exchanges at metropolitan distances, with a locally generated \glsxtrlong{lo}, frequency multiplexed pilots, using the RF-heterodyne technique for the detection. We report here on asymptotic key rates up to $22\;\si{\mega\bit\per\second}$ at zero distance and $1.2\;\si{\mega\bit\per\second}$ at $25\;\si{\kilo\meter}$ achieved using our open source platform. The extension of its use to other \gls{cvqkd} setups has either been investigated or should be possible without high complexity. 

The rest of this paper is structured as follows: in section~\ref{sec:protocol-setup}, we describe the protocol and the chosen setup, while in section~\ref{sec:qosst}, we present our open source software for \gls{cvqkd}, QOSST, along with an in-depth explanation of the \gls{dsp} algorithm. In section~\ref{sec:results}, an experimental benchmark of the software using our optical setup is presented, before discussing the results and drawing conclusions in sections~\ref{sec:discussion} and \ref{sec:conclusion} respectively.

\section{Protocol and setup\label{sec:protocol-setup}}

\subsection{Description of the protocol}

\begin{figure*}
    \centering
    \includegraphics[width=0.9\textwidth]{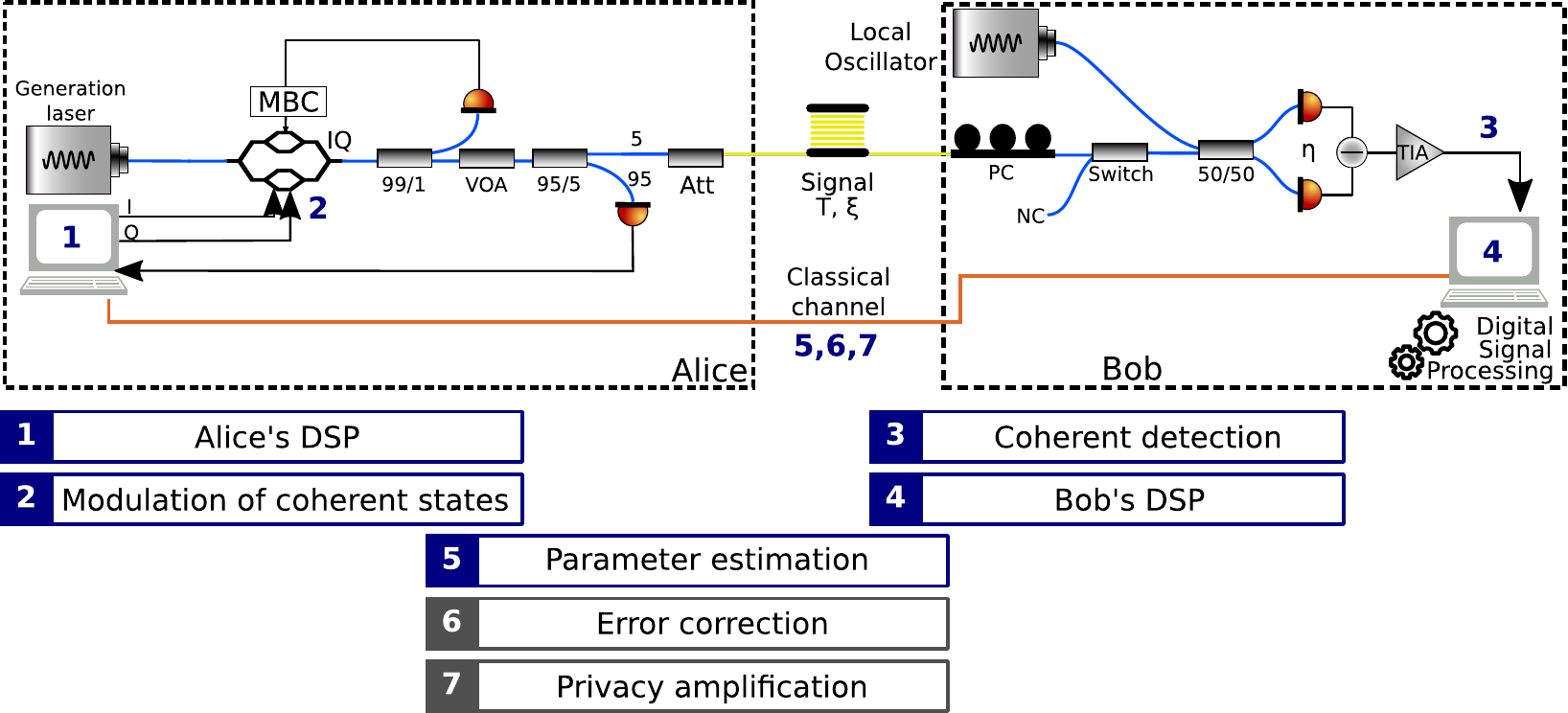}
    \caption{Scheme of the CV-QKD optical setup and general steps of the protocol. Here we show a fiber spool but a \glsxtrlong{voa} can also be used as a channel. In this case, the polarisation controller at the input of Bob is not present. The attenuator at Alice's side introduces a fixed attenuation of $10\;\si{dB}$. 
    Acronyms: IQ: IQ modulator, MBC: Modulator Bias Controller, VOA: Variable Optical Attenuator, Att: Attenuator (fixed), PC: Polarisation controller, NC: Not Connected, TIA: Trans-Impedance Amplifier, DSP: Digital Signal Processing. The details of the setup are provided in the main text.}
    \label{fig:cvqkd_schema}
\end{figure*}

In general terms, \gls{cvqkd} protocols follow the steps shown in Fig.~\ref{fig:cvqkd_schema}. Alice randomly chooses symbols drawn from a specific constellation, encodes them on quantum states and sends them to Bob through a quantum channel, which is characterised by a transmittance $T$ and an excess noise $\xi$. Bob then measures the quantum states using coherent detection techniques. In some protocols, Bob needs to randomly choose which quadrature he measures at this moment. Then, Bob performs his \gls{dsp} to recover the symbols and estimate the parameters of the channel ($T$ and $\xi$) such that he can upper bound Eve's information, assuming that all the excess noise comes from eavesdropping. Finally, Alice and Bob perform error correction and privacy amplification. These last two steps are not implemented in the present work.

We now discuss in more detail the protocol steps. Alice's symbols are defined by complex numbers, whose real and imaginary parts are drawn from a particular probability distribution. CV-QKD systems up until recently have typically used Gaussian modulation, meaning that the real and imaginary parts of the symbols follow a Gaussian distribution. However, more recently discrete modulation was proposed as an attractive alternative to align with standard practices in coherent optical communications and optimise bandwidth use and classical post-processing. In this case, the set of possible symbols (the constellation) is finite. For instance, the $M$-\gls{psk} constellation has $M$ possible symbols with the same amplitude (the symbols are spaced on a circle with an angle difference $2\pi/M$). The $M$-\gls{qam}, with $M$ being a perfect square, is also a discrete modulation where the symbols are placed on a $\sqrt{M}\times\sqrt{M}$ grid. It is also possible to approximate a Gaussian distribution with a \gls{qam}, using \gls{pcs} or a Binomial Distribution. Examples of such modulations are shown in Fig.~\ref{fig:modulations}.

Security proofs for Gaussian modulation are well established, including also finite-size effects. Despite the complexity in proving security for discrete constellations, a security proof is now also available for any such constellation at the asymptotic limit~\cite{ExplicitAsymptDenys2021}, and was used in a recent implementation with PCS-QAM modulation~\cite{ExperimentalDeRoumes2024}.


In our work here, we use as baseline protocol the \gls{gmcs} GG02 protocol~\cite{ContinuousVariGrossh2002}. After the generation of the symbols on Alice's side with using a random number generator (ideally a true random number generator such as a \gls{qrng}), we prepare them for transmission applying techniques inspired by classical communications. In particular, the quantum symbols are filtered using a \gls{rrc} filter to occupy a finite bandwidth $B = R_s\cdot (1+\beta_\mathrm{RRC})$ where $R_s$ is the symbol rate and $\beta_\mathrm{RRC}$ is the so-called roll-off factor (taking values between 0 and 1). The \gls{rrc} filter is also applied on Bob's side, giving, at the end, a \gls{rc} filter, which fulfills the Nyquist criteria to minimise inter-symbol interference~\cite{Proakis2007}.

In addition, the DSP algorithm involves the use of classical signals to perform frame synchronisation, clock recovery, carrier frequency estimation and phase compensation. For the synchronisation, we use a \gls{cazac} sequence, namely a Zadoff-Chu sequence~\cite{PhaseShiftPulHeimil1961, PolyphaseCodesChuD1972, Zepernick2005-vn}. This sequence has good correlation properties that allow for an easy synchronisation and is added before the quantum data. For clock recovery, carrier frequency estimation and phase compensation, we use two frequency multiplexed pilot tones, which are complex exponentials at defined frequencies. More details on the \gls{rc} filter and Zadoff-Chu sequence can be found in appendix~\ref{appendix:dsp}.

The signal is modulated in one optical sideband (\gls{ossb} modulation) and measured by Bob using the RF-heterodyne technique (see appendix~\ref{appendix:rf-heterodyne} for more details). We refer to the whole protocol as \gls{ossb}-\gls{gmcs}-GG02.

Once Bob's \gls{dsp} is performed, Alice and Bob proceed to the parameter estimation step. Denoting as $X$ and $Y$ the random variables describing the symbols of Alice and Bob, respectively, we have:
\begin{equation}
    Y = \sqrt{\eta T}X + N.
\end{equation}
Here $N$ is the white Gaussian noise with variance $1+V_{el}+\eta T \xi$ in Shot Noise Units (SNU), where $V_{el}$ and $\eta$ are the electronic noise and quantum efficiency of the detector, respectively.  

Then, if $V_A$ is the modulation variance at Alice's side, the following formulas hold:
\begin{equation}
    \begin{split}
        \langle X^2 \rangle &= V_A\\
        \langle XY \rangle &= \sqrt{\frac{\eta T}{2}}V_A\\
        \langle Y^2 \rangle &= 1 + V_{el} + \frac{\eta T}{2} V_A + \frac{\eta T}{2} \xi
    \end{split}
\end{equation}
and hence, if $\eta$, $V_{el}$ and $V_A$ are known, we have
\begin{equation}
    \begin{split}
        T &= 2\cdot\frac{\langle XY\rangle^2}{V_A^2 \eta}\\
        \xi &= 2\cdot\frac{\langle Y^2 \rangle - 1 - V_{el} - \frac{\eta T}{2} V_A}{\eta T}
    \end{split}
    \label{eqn:parameters_estimation}
\end{equation}
We can also define the excess noise at Bob's side as $\xi_B = \eta T \xi/2$ or $\xi_B = \langle Y^2 \rangle - 1 - V_{el} - \eta T V_A/2$.

After parameter estimation, according to the asymptotic security proof for Gaussian modulation and for reverse reconciliation where Bob's data is used to form the key, it is possible to bound the information that Eve has acquired using the Holevo bound $\chi_{BE}$ and compute the \gls{skr} using the Devetak-Winter formula~\cite{DistillationOfDeveta2005}:
\begin{equation}
    K_f = \beta I_{AB}(V_A, T, \xi, \eta, V_{el}) - \chi_{BE}(V_A, T, \xi, \eta, V_{el}),
    \label{eqn:skr}
\end{equation}
where $\beta$ is the reconciliation efficiency and $I_{AB}$ is the mutual information shared by Alice and Bob. In the security analysis, we make the standard assumption in CV-QKD implementations of a trusted detector, which means that Eve cannot interfere with Bob's receiver. This allows us to separate the losses ($\eta$) and the electronic noise ($V_{el}$) of the receiver and to consider only the remaining losses ($T$) and noise ($\xi$) as a signature of the presence of Eve, giving a more favourable bound on the leaked information.

To take into account the finite number of samples used to estimate the values of $T$ and $\xi$ in practice, we use worst-case estimators as follows: assuming that $\xi_B$ and $T$ are sampled from a Gaussian probability distribution, we compute the values $x_B^w$ and $T^w$ such that the probability that the actual value of $\xi_B$ (resp. $T$) is above $\xi_B^w$ (resp. below $T^w$) is $\varepsilon$, which in our case  is chosen to be $\varepsilon=10^{-10}$. We also perform finite-size analysis following the analysis of~\cite{FiniteSizeAnaLeverr2010} (see section~\ref{sec:discussion} for more details).

The key rate that we compute in this way gives us the number of secret bits per symbol. To get the secret key rate in bits/s, one can apply the formula $K_s = R_s\cdot K_f$, where $R_s$ is the symbol rate.

Finally, in the complete protocol Alice and Bob perform error correction and privacy amplification. Error correction can be done for instance using techniques described in~\cite{HighSpeedErroWang2018,ANovelErrorCGumus2021}. Privacy amplification compresses the key so that Eve does not hold any meaningful information on it, and is typically done using Toeplitz matrices~\cite{MemorySavingABaiE2022}. These steps are not implemented in QOSST yet and are beyond the scope of this paper.

\subsection{Experimental setup}

The experimental setup is shown in Fig.~\ref{fig:cvqkd_schema} together with the protocol steps. The setup for generating the quantum states at Alice's side is composed of a continuous wave laser (NKT Koheras Basik), an IQ modulator (Exail MXIQER-LN-30), a \gls{mbc} (Exail MBC-IQ-LAB), an electronic \gls{voa} (Thorlabs V1550PA), followed by a 95:5 beam splitter, with the 95\% output going into the monitoring photodiode (Thorlabs PM101A) and the 5\% output going into a 90:10 beam splitter, whose 10\% output is sent to the output of Alice. Alice is also composed of a control PC and a \gls{dac} (Teledyne SDR14Tx). The rate of the \gls{dac} is $2\;\si{\giga\sample\per\second}$ with a bandwidth of $1\;\si{\giga\hertz}$, and it has a 14-bit resolution. The \gls{dac} is connected to the control PC using a PCIe connector.

The role of the \gls{mbc} is to apply the correct bias DC voltages to the modulator using a feedback loop. Indeed, a single polarization IQ modulator requires 3 bias voltages that can be found by adding low frequency dithers into the DC input and applying an optimization algorithm. In this way, it is possible to maximise the extinction ratio and to adjust the relative phase between the two quadratures to $90^\circ$. The role of the monitoring photodiode is to measure the number of photons per symbol in the quantum data: if the ratio $r_{\mathrm{conv}}$ between the output power of Alice and the power at the monitoring photodiode is known (see appendix~\ref{appendix:calibration-alice} for more details on the calibration of Alice), then the average number of photons per symbol at Alice's output is given by
\begin{equation}
    \langle n \rangle = \frac{r_{\mathrm{conv}}\cdot P_\mathrm{monitoring}}{E_{\mathrm{ph}}\cdot R_s},
\end{equation}
where $R_s$ is the symbol rate, $P_\mathrm{monitoring}$ is the optical power on the monitoring photodiode for the quantum data, and $E_{\mathrm{ph}} = h c/\lambda$ is the energy of the photon. The average number of photons per symbol allows us to compute Alice's variance $V_A$ using the following formula:
\begin{equation}
    V_A = 2\cdot\langle n \rangle.
\end{equation}
The final 90:10 beam splitter is used to provide a further 10 dB attenuation. The electronic \gls{voa} is used, together with the power of the laser and the state generation parameters of Alice's \gls{dsp}, to tune both Alice's variance and the power of the classical signals (pilot tones). It is controlled using a National Instruments card (USB-6363).

The entrance of the setup at Bob's side is controlled by an optical switch (Thorlabs OSW12-1310E), which can be used to disconnect the quantum channel from the detection system during the calibration, in order to prevent an eavesdropper  from tampering with this procedure. This also ensures that noise coming from Alice's setup is not inadvertently included in the calibration. The output of the switch goes into a 50:50 beam splitter, where the signal is mixed with a continuous wave laser (NKT Koheras Basik) acting as a \gls{llo}. The two outputs of the 50:50 beam splitter go into a \gls{bhd} (Thorlabs PDB480C-AC), whose output is filtered using an analog low-pass filter with a cut-off frequency of $700\;\si{\mega\hertz}$ before being sent to Bob's control PC through an \gls{adc} (Teledyne ADQ32) working at $2.5\;\si{\giga\sample\per\second}$ and with a bandwidth of $1.5\;\si{\giga\hertz}$. The \gls{adc} is connected to the PC using a PCIe connector. Bob's setup also requires calibration, in particular for the efficiency and the electronic noise (see appendix~\ref{appendix:calibration-bob} for details).

Both the transmitter setup of Alice and the receiver setup of Bob use \gls{pm} components.
The quantum channel between Alice and Bob in our experiments is implemented either with a \gls{pm} electronic \gls{voa} (Thorlabs V1550PA) or with a single mode fiber spool with a length of $25.2\;\si{\kilo\meter}$ (linear losses: $4.75\;\si{\deci\bel}$, attenuation coefficient: $0.188\;\si{\deci\bel\per\kilo\meter}$, total losses with connector: $5.22 \;\si{\deci\bel}$), whose characteristics are similar to the fibers deployed in standard optical fiber networks. When using a single mode fiber spool, a manual polarisation controller placed at Bob's input, before the optical switch, is used to recover the polarisation after the transmission.

Both control PCs are connected to the same local network to provide the classical channel. Authentication on the classical channel is done using the Post Quantum Algorithm for Digital Signatures Falcon~\cite{falcon,falcon-implementation-python,falcon-implementation-python-yoann}, which is expected to be standardised in the NIST Post Quantum Cryptography standardization process~\cite{nist-round-4}.
There is no other link between Alice and Bob.

\section{Description of the open source platform\label{sec:qosst}}

We now present our software, called QOSST (\textit{Quantum Open Software for Secure Transmissions}), that was developed in our team and aims at performing all basic operations that are necessary for \gls{cvqkd} implementations.

\subsection{Design of the software}

The software is written using the Python language~\cite{python}, and was conceived to be modular. Importantly, it is in principle not limited to a specific hardware and transmission or detection schemes. It might however require some work to adapt the \gls{dsp} for operations with time-multiplexed pilots or in the pulsed regime. 

QOSST is released as an open source software: the different repositories can be found at \url{https://github.com/qosst} and a good starting point to understand QOSST would be \url{https://github.com/qosst/qosst}.

\begin{figure}
    \centering
    \includegraphics[scale=0.6]{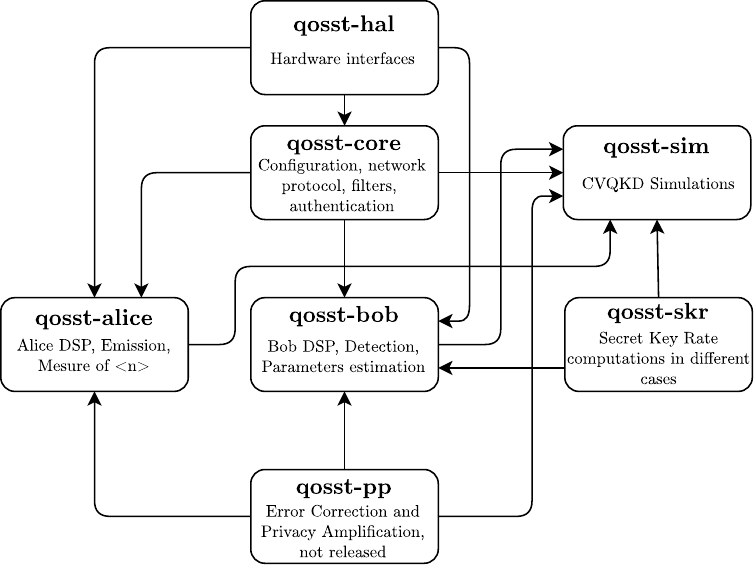}
    \caption{Interconnections of the QOSST packages along with the main functionalities. An arrow from block A to block B means that the block A is a dependence of block B. For instance, the arrow from qosst-core to qosst-alice means that qosst-alice uses qosst-core. The details of the packages are provided in the main text.}
    \label{fig:qosst-interconnections}
\end{figure}

The software is separated in 7 packages, out of which 6 are released and 1 is still under development and has not been released yet. The structure is shown in Fig.~\ref{fig:qosst-interconnections} along with the interconnections between the packages and the main functionalities. In the following, we provide details on these packages.

\subsection{Hardware Abstraction Layer}

The \gls{hal} package provides abstract interface classes for the hardware. In this sense, it is easy to write a class to control a new hardware and then replace the class in the configuration file. This operation should be totally transparent for the rest of QOSST. This package also provides dummy classes that can be used as default, and special values when the hardware is not used. The classes for the actual hardware that we used are not released with QOSST. A tutorial on how to write new classes is available in the documentation.

\subsection{Core}

The core package of QOSST is the biggest of the seven and implements the functions that are common to Alice and Bob. In particular, it implements the control protocol, the configuration, the filters, the synchronisation sequence, the modulations, the authentication, the data containers and some common functions.

A specific network protocol for the classical channel has been designed for QOSST and its specifications are described in the documentation. It has a built-in authentication system that can be used to authenticate the classical channel. The package also provides a socket class to easily interact between a server and a client.

The configuration file, which controls the whole protocol, is written in TOML~\cite{toml} since the Python ecosystem appears to adopt TOML as its default configuration language~\cite{pep518,pep680} and this language is also easily readable and modifiable by both humans and machines~\cite{pep518,comparison-file-languages}. The configuration holds more than 100 parameters and part of them needs to be optimized to reach the best values of excess noise and secret key rate. An example configuration file is provided with the software and the explanation of all the parameters can be found in the documentation.

The package also implements the Zadoff-Chu sequence for synchronisation and the following filters: \gls{rc}, \gls{rrc} and pulsed (filter with a rectangular temporal response).

\begin{figure*}
    \centering
    \includegraphics{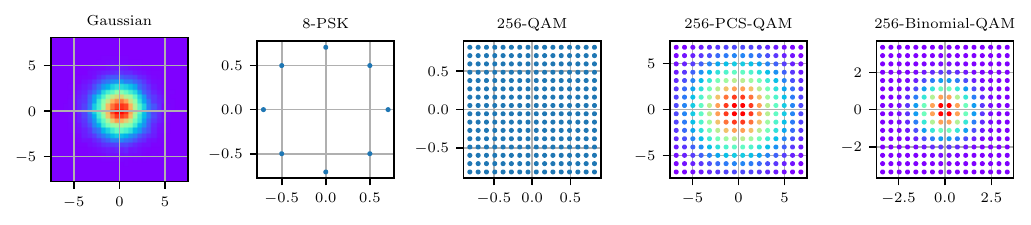}
    \caption{Possible modulation schemes in the QOSST software. From left to right: Gaussian (variance: 3, number of points: 100000), 8-PSK (variance: 1), 256-QAM (variance: 1), 256-PCS-QAM (variance: 35, $\nu=0.01$~\cite{ExperimentalDeRoumes2024}), 256-Binomial-QAM (variance: 3). The variance parameter does not always represent the actual variance of the generated sequence, but is a parameter that is proportional to the actual variance.}
    \label{fig:modulations}
\end{figure*}

The package provides the codes for different modulation schemes, which can be used instead of the default Gaussian modulation: $M$-\gls{psk}, $M$-\gls{qam}, \gls{pcs}-\gls{qam}, Binomial \gls{qam}. The modulations are shown in Fig.~\ref{fig:modulations}.

The authentication is handled by adding a signed digest in the control protocol frame. In the QOSST implementation, the digest can either be not signed or signed with the Falcon algorithm.

We also note that the documentation includes the description of some other useful functions present in QOSST but that are not relevant for this paper.

\subsection{Alice}

The Alice package implements mainly two functionalities: the first one is the server, with which Bob's client interacts, and the other one is the \gls{dsp} for signal generation.
The server answers to valid Bob's requests. In particular, it generates the data and applies it to the IQ modulator when asked to do so.

\begin{figure}
    \centering
    \includegraphics{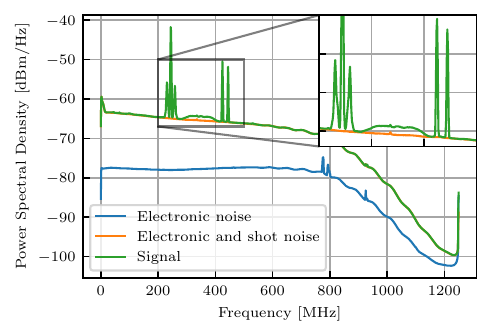}
    \caption{Power Spectral Densities of the electronic noise, electronic and shot noise and signal data. In this example, we can see that $f_\mathrm{beat}$ is around $240\;\si{\mega\hertz}$, and that at this frequency, there is some ``low-frequency'' noise coming from the setup of Alice. This does not impact the performance of the \gls{cvqkd} protocol since it is filtered out. The quantum signal has a bandwidth of $140\;\si{\mega\hertz}$ and is centered at $f_\mathrm{beat}+f_\mathrm{shift}$ with $f_\mathrm{shift}=100\;\si{\mega\hertz}$. The two pilots are placed at the frequencies $f_\mathrm{beat}+f_{\mathrm{pilot},1}$ and $f_\mathrm{beat}+f_{\mathrm{pilot},2}$ with $f_{\mathrm{pilot}, 1} = 180\;\si{\mega\hertz}$ and $f_{\mathrm{pilot}, 2} = 200\;\si{\mega\hertz}$. The first pilot is slightly more powerful than the second one to be consistent when recovering the frequencies of the pilots. The slight signal at $f_\mathrm{beat}-f_\mathrm{shift}$, $f_\mathrm{beat}-f_{\mathrm{pilot},1}$ and $f_\mathrm{beat}-f_{pilot,2}$ is due to imperfections on the sideband suppression.}
    \label{fig:frequential}
\end{figure}

In the following, we describe the \gls{dsp} algorithm for signal generation using Nyquist filters and frequency multiplexed pilots. The first step is the generation of the random symbols from the chosen distribution~\footnote{The symbols can also be loaded from a file.}. In QOSST, at the time of writing of this paper, the entropy source is a pseudo random number generator from the \texttt{numpy} library~\cite{numpy}; in practice, entropy should be provided by a \gls{qrng} and this can be easily integrated in QOSST. The symbols are then upsampled to match the symbol rate $R_s$ and filtered using a \gls{rrc} cosine filter with roll-off factor $\beta_\mathrm{RRC}$. The quantum data is then shifted in frequency to the center frequency $f_\mathrm{shift}$. Two pilot tones, of frequencies $f_{\mathrm{pilot},1}$ and $f_{\mathrm{pilot},2}$, are frequency multiplexed with the signal. Their frequencies are chosen so that
\begin{equation}
    f_{\mathrm{pilot},1}, f_{\mathrm{pilot},2} \notin \left[f_\mathrm{shift} - \frac{B_s}{2}, f_\mathrm{shift} + \frac{B_s}{2}\right],
    \label{eqn:pilots_constraint}
\end{equation}
with
\begin{equation}
    B_s = R_s\cdot (1+\beta_\mathrm{RRC})
    \label{eqn:bandwidth_rrc}
\end{equation}
the bandwidth of the quantum data. Then a Zadoff-Chu sequence with length $L_{ZC}$ and root $R_{ZC}$ is generated and prepended to the signal. The \gls{dsp} can also prepend and append a sequence of zeros if needed. The frequency form of the signal is shown in Fig.~\ref{fig:frequential}.

Alice's server applies this sequence to the IQ modulator when Bob sends the network request indicating he is ready. Alice's server also estimates the average number of photons per symbol $\langle n \rangle$ and sends the data for the parameter estimation to Bob.
Finally, this package provides a script to calibrate the conversion factor $r_{\mathrm{conv}}$.

\subsection{Bob}

This is the second biggest package of QOSST and provides the \gls{dsp} and scripts for Bob.
First, the package provides ways to characterize the receiver, in particular to measure the detector efficiency $\eta$ and the electronic noise $V_{el}$ (more precisely, the sequence of electronic noise samples that can be used to compute $V_{el}$ afterwards).
Then, Bob provides a client that can either be used with command line scripts or as a \gls{gui}, which interacts with the server of Alice. In particular, the client starts the acquisition, triggers Alice, and then applies the \gls{dsp} to recover the symbols. The client also requests Alice's symbols and performs the parameter estimation.

Again, we give an overview of the \gls{dsp} algorithm for our protocol. First, if the automatic shot noise calibration is enabled, Bob uses the beginning of the data (before the switch connects his setup with the quantum channel) to estimate the shot noise value, denoted as $\sigma_0^2$, that will be used for normalisation of the symbols in SNU at Bob's side (more precisely, the sequence of electronic and shot noise samples that can be used to estimate $\sigma_0^2$ afterwards, as will be explained later). Then, the first step of the actual \gls{dsp} is to find an estimate of the position of the Zadoff-Chu sequence. Knowing this information, it is possible to isolate the part of the data with the pilots and to search in the frequency domain the two frequencies of the pilots: $\tilde{f}_{\mathrm{pilot}, 1}^B$ and $\tilde{f}_{\mathrm{pilot}, 2}^B$. The clock difference is estimated using the formula
\begin{equation}
    \Delta f = \frac{\tilde{f}_{\mathrm{pilot}, 2}^B - \tilde{f}_{\mathrm{pilot}, 1}^B}{f_{\mathrm{pilot}, 2}-f_{\mathrm{pilot}, 1}},
\end{equation}
which is then used to correct the clock mismatch. This step is known as clock recovery. Once the clock is recovered, the frequencies of the pilots are estimated again to find $f_{\mathrm{pilot}, 1}^B$ and $f_{\mathrm{pilot}, 2}^B$ and the beat frequency is computed as
\begin{equation}
    f_\mathrm{beat} = f_{\mathrm{pilot}, 1}^B - f_{\mathrm{pilot}, 1}.
    \label{eqn:fbeat}
\end{equation}

The beat frequency is the frequency difference between the two lasers, and can typically move by a few $\si{\mega\hertz}$ if the lasers are not very stable. This was not the case in our implementation thanks to the stability of our lasers. The sequence is unshifted by $f_\mathrm{beat}$. This step is known as carrier frequency estimation. Then, the Zadoff-Chu sequence is found by cross-correlations, which gives a precise value for the beginning of the sequence. Knowing the number of symbols and the symbol rate, the sequence is reduced to the useful sequence containing only the symbols. The rest of the \gls{dsp} is performed by subframes because the change in $f_\mathrm{beat}$ can be relatively quick and we want to do the analysis on a time scale where $f_\mathrm{beat}$ can be considered constant. In particular, this subframe analysis is critical when the lasers used are not very stable. {In particular the duration of the subframe must be chosen such that the relative frequency variations between the two lasers (including linewidth and change in central frequencies) are less than the invert of the subframe's duration, during the said subframe. In practice, this can be done by an optimization over the possible subframe sizes as shown in the middle graph of Fig.~\ref{fig:optimisations}.} For each subframe, the frequencies of the two pilots are estimated again, and $f_\mathrm{beat}$ is also estimated again using Eq.~(\ref{eqn:fbeat}). The first pilot is filtered and used as a phase reference, and a matching \gls{rrc} filter is applied to the quantum data, after it has been shifted to baseband by $f_\mathrm{beat}+f_\mathrm{shift}$. Following this, the best sampling point is found by the maximal variance (eye diagram) method~\cite{7174950}. The phase is then corrected using the phase reference from the pilot (a uniform filter is applied to the phase reference for better performance). At this point, Alice's and Bob's symbols only differ by a global phase (and some noise) and the global phase correction is performed by testing all global rotations and keeping the one with the best covariance term $\langle XY \rangle$. More details on the \gls{dsp} of both Alice and Bob can be found in appendix~\ref{appendix:dsp}.

After these steps, Bob applies the same \gls{dsp} to the electronic noise and electronic and shot noise samples, whose outputs give the electronic noise and electronic and shot noise symbols. The program also requests $\langle n \rangle$ from Alice. Bob then has the electronic noise symbols, the electronic and shot noise symbols, the average number of photons per symbol $\langle n \rangle$, Alice's symbols (for estimation) and Bob's symbols. The variance of the electronic noise symbols is $\sigma_{el}^2$ and the variance of the electronic and shot noise symbols is $\sigma_{el}^2+\sigma_{0}^2$, which allows to calculate $\sigma_{0}^2$. $V_{el}$ is then calculated as $V_{el} = \sigma_{el}^2/\sigma_0^2$ and $T$ and $\xi$ can be calculated with the formulas in Eq.~(\ref{eqn:parameters_estimation}) using the appropriate normalisation (see appendix~\ref{appendix:snu}
for more details).

In this way, Bob ends up with $T$, $\xi$, $\langle n \rangle$, $\eta$ and $V_{el}$ and can proceed to the estimation of the secret key rate using Eq.~(\ref{eqn:skr}) (using code from the skr package). We typically use the value $\beta=0.95$ for the reconciliation efficiency~\cite{LongDistanceCJougue2011}. 

The Bob package also provides automated scripts to allow for performing the experiment over long periods of time.

\subsection{Secret Key Rate}

The skr package provides classes with static methods to compute the secret key rate depending on the setup and the security assumptions. Three methods are currently available: untrusted homodyne detector~\cite{leverrier:tel-00451021}, trusted homodyne detector~\cite{QuantumKeyDisLodewy2007} and trusted heterodyne detector~\cite{Fossier_2009}, all in the asymptotic case. In particular, the code that is released in this package does not take into account discrete modulations and finite-size effects.

{Technically, QOSST can emit and receive discrete-modulated sequences (although the DSP does not use in its present form any potential advantage provided by the discrete nature of the constellation). However, QOSST does not yet include methods to compute the SKR in the discrete-modulation scenario. We have planned to add, in future releases of this package, methods following the analysis of Denys and Leverrier~\cite{ExplicitAsymptDenys2021} for the asymptotic case. The finite-size case is more delicate as, while much progress has been made in the past few years, it is still unclear which analysis will provide the best key rates in practice.}

\subsection{Simulations and Post-Processing}

The last two packages of QOSST are sim (simulations) and pp (Post Processing). The first one, qosst-sim, is also released in the software suite and allows for the simulation of \gls{cvqkd} exchanges.

While a partial codebase for the post processing package exists, including error correction with LDPC codes, and privacy amplification with Toeplitz matrices, its content was judged not mature enough for publication.

{
\subsection{Re-usability of the software}

In the hope that our open source platform can be of use to the community, we have undertaken several steps to enhance the re-usability of our code. First, the Hardware Abstraction Layer allows to use the code by only providing the new control commands, and changing a parameter in the configuration file, hence allowing for the use of other pieces of equipment with minimal changes. Second, the behavior of the software is highly configurable with more than 100 parameters in the configuration file, allowing for the adaptation to a variety of situations and optimizations. Finally, the code has been broken down into relatively small functions, allowing their use outside of the main program. For instance, it is easy to use the functions for the DSP, for parameter estimation or for key rate computations, without having to run the full program or use any hardware.

Our software has, however, some limitations in its present form: it cannot be directly used to experiment with time-multiplexed pilots, phase-diverse detection or polarization-diverse setups. While we plan to provide updates to increase the scope of our software in future releases, we hope that the provided documentation is already sufficient for other groups to adapt our software for their experiment, if needed.

}

\section{Results\label{sec:results}}

To assess the performance of CV-QKD driven by QOSST, we tested and benchmarked it using the experimental setup presented in section~\ref{sec:protocol-setup}. For this, we used five channel configurations. In four of them, the \gls{pm} \gls{voa} emulates an optical fiber of $0$, $5$, $10$ and $25\;\si{\kilo\meter}$ at $0.2\;\si{\deci\bel\per\kilo\meter}$ (the actual attenuation values were in practice slightly higher). The last configuration uses a fiber spool of $25\;\si{\kilo\meter}$ (total attenuation: $5.22\;\si{\deci\bel}$) and a manual polarisation controller to compensate for the polarisation transformation of the fiber.

For each configuration, we optimized around 10 \gls{dsp} and physical parameters using automated scripts, including the \gls{rrc} filter's roll-off factor, the pilot amplitudes, the average number of photons $\langle n \rangle$, the power of Alice's laser, the power of Bob's laser, the size of the subframes, the width of the filters for the tones, the length of the uniform filter of the phase, the baud rate, the frequency shift and the frequency difference between the pilots. The optimization is done automatically using the \texttt{qosst-bob-optimize} script, which exchanges frames and dynamically changes the parameters of Alice and Bob.

\begin{figure*}
    \centering
    \includegraphics{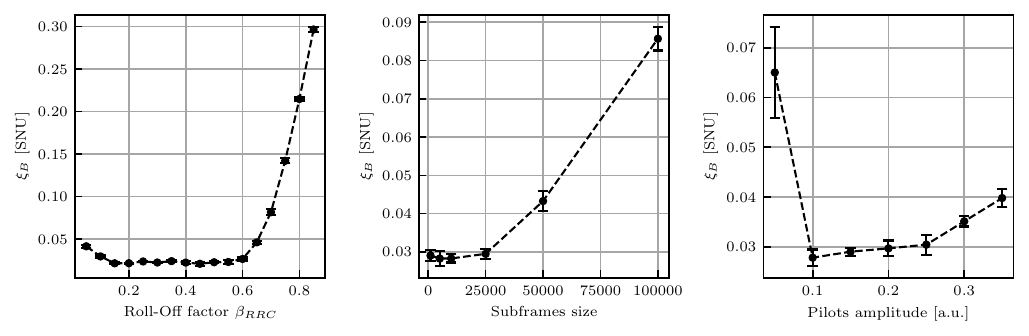}
    \caption{Results of three examples of optimisation. For the three experiments, the number of frames per point is 5. For the pilot amplitude a fine search was conducted after a first coarse one to find the optimal value in the 0.05-0.15 region.}
    \label{fig:optimisations}
\end{figure*}

Here we give examples of optimization of three parameters in terms of the excess noise at Bob's side, even though it is required to perform the optimization on all the parameters to reach the best performance.
For each point on each experiment, five frames were exchanged and the results are given in Fig.~\ref{fig:optimisations}.

For the roll-off factor, we see that a value below $0.2$ induces a slightly higher excess noise, and for high values a steep increase of the excess noise is observed. We can explain this effect by recalling that the bandwidth of the signal after the \gls{rrc} filter is given by Eq.~(\ref{eqn:bandwidth_rrc}). Since in this experiment the frequencies of the pilots are fixed, Eq.~(\ref{eqn:pilots_constraint}) is violated when $\beta_\mathrm{RRC}$ becomes too big. In practice, the quantum data and the pilots start to be superposed in frequency, increasing the noise on the quantum data and reducing the precision of the phase compensation algorithm. This effect starts when $\beta_\mathrm{RRC}=0.65$, which corresponds to having $f_\mathrm{shift} + B_s/2 = 182.5\;\si{\mega\hertz}$ and is the first roll-off value where $f_\mathrm{shift} + B_s/2 > f_{\mathrm{pilot},1} = 180\;\si{\mega\hertz}$.

For the subframe size, we see that the excess noise tends to increase when the size increases. This is reasonable as the bigger the subframe, the more the $f_\mathrm{beat}$ frequency is moving and the less precise the \gls{dsp} algorithm is. However, for very small values of subframe size, the excess noise also seems to be slightly higher. This was attributed to the fact that, in this regime, there are not enough samples to properly estimate the frequencies of the pilots.

Finally for the pilot amplitudes, a too low value results in a very high excess noise. This is due to the fact that either the frequency of the pilots cannot be properly recovered, or the \gls{snr} of the pilots is too low to get a good phase compensation. Also, the excess noise increases with the pilot amplitudes, which can be explained by frequency cross-talk between the quantum data and the pilots. The optimal value depends mainly on the \gls{snr} at the receiver, which means that this value should be optimized for every distance. This optimization was done for each of the 5 configurations. This first coarse optimisation was also followed by a fine optimisation (in this example in the region 0.05-0.15). The amplitude of the pilots is given in arbitrary units, and the actual amplitude of the pilot is proportional to this value. The optimal value was found to correspond to $\sim12\;\si{\deci\bel}$. 

Our tests showed that the optimal value of some parameters is independent of the attenuation. The values of these parameters are given in table~\ref{tab:params-ind-distance}, along with some other important fixed parameters in our experiments.
\begin{table}
    \centering
    \begin{tabular}{c|c}
        \textbf{Parameter} & \textbf{Value}                      \\
        \hline
        $\beta_\mathrm{RRC}$      & $0.5$                               \\
        $R_s$              & $100\;\si{\mega\baud}$              \\
        $f_\mathrm{shift}$        & $100\;\si{\mega\hertz}$             \\
        $f_{\mathrm{pilot}, 1}$     & $180\;\si{\mega\hertz}$             \\
        $f_{\mathrm{pilot}, 2}$     & $200\;\si{\mega\hertz}$             \\
        $L_{ZC}$           & $3989$                              \\
        $R_{ZC}$           & $5$                                 \\
        Acquisition time   & $50\;\si{\milli\second}$            \\
        Shot noise time    & $10\;\si{\milli\second}$            \\
        DAC rate           & $2\;\si{\giga\sample\per\second}$ \\
        ADC rate           & $2.5\;\si{\giga\sample\per\second}$ \\
        Modulation         & Gaussian                            \\
    \end{tabular}
    \caption{Values of the main parameters that are independent of the distance and other fixed parameters in our setup.}
    \label{tab:params-ind-distance}
\end{table}
Once all the parameters were optimized, an experiment of 200 \gls{cvqkd} frames was launched for each configuration, which corresponds to around 10 to 12 hours. In our setup, the instability, due for instance to mechanical vibrations, was not compensated. This does not necessarily lead to negative key rates for some frames but it may affect the worst-case estimators in the finite-size analysis because of an increased variance, leading to pessimistic values. 
Especially for the fiber experiment, the effect of instability is important as, after some time, the polarisation mismatch becomes too big and the efficiency of the detection drops, giving a null key rate.

\begin{figure*}
    \centering
    \includegraphics[width=0.9\textwidth]{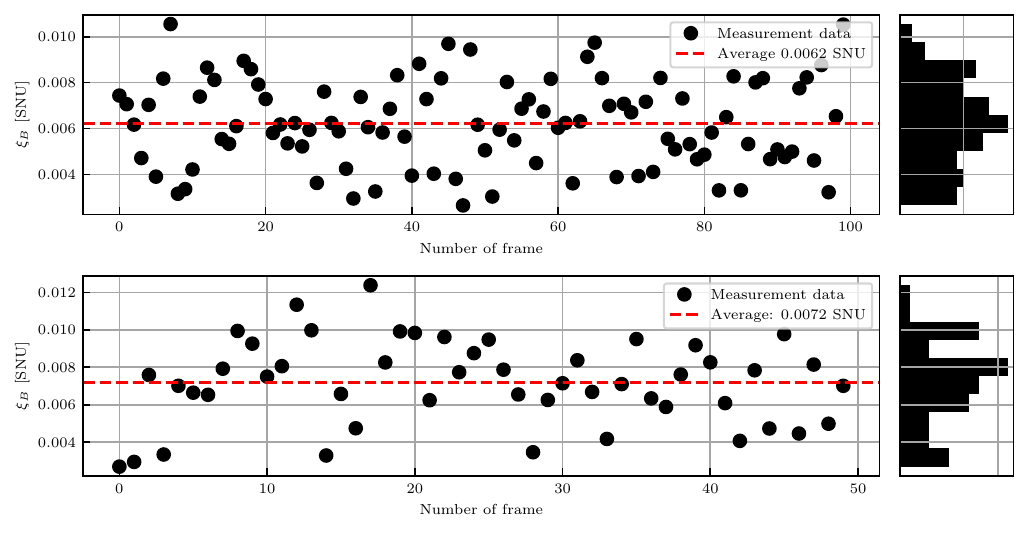}
    \caption{Excess noise at Bob's side for the experiments with the VOA (equivalent to 25~km) and the 25~km fiber spool. Top: experiment with the VOA. One frame where the \gls{dsp} failed was removed from the 100 frames. The average excess noise at Bob's side is $0.0062\;\si{\snu}$. Bottom: experiment with the fiber. Only the first 50 frames where the experiment was stable are shown. During this time, the average excess noise at Bob's side is $0.0072\;\si{\snu}$.}
    \label{fig:excess_noise}
\end{figure*}

\begin{figure}
    \centering
    \includegraphics{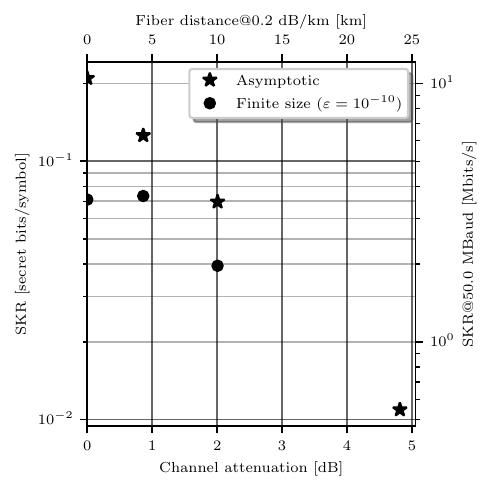}
    \caption{Achievable secret key rate in the asymptotic and finite-size regime. The security parameter is $\varepsilon = 10^{-10}$. For each experiment, the actual attenuation performed by the \gls{voa}  was always slightly above the targeted one. For the experiments with the \gls{voa}, the extra attenuation corresponds to a slightly higher attenuation coefficient on the actual fiber, and a small mismatch in polarisation that is not included in the efficiency. For the experiment with the fiber spool and the \gls{voa} at $25\;\si{\kilo\meter}$ the finite-size secret key rate was zero.}
    \label{fig:skr}
\end{figure}

In Fig.~\ref{fig:excess_noise}, we plot the excess noise for the experiments using the \gls{voa} with attenuation corresponding to $25\;\si{\kilo\meter}$ and the fiber spool. The estimation of the excess noise is rather stable. The stability of the estimation of $\langle n \rangle$, $T$, shot noise and $V_{el}$ were also checked. The results in terms of the secret key rate are given in Fig.~\ref{fig:skr}. {Note that for the fiber spool experiment, only the first 50 frames are shown. Indeed, after this time, the setup drifts away from its optimal position, in particular the polarization of the signal and of the local oscillator start to mismatch, causing a drop in efficiency. This technical issue has now been resolved, as discussed later.}

\begin{table}
    \centering
    \scalebox{0.95}{\begin{tabular}{c|c|c|c}
        \multirow{2}{*}{\textbf{Experiment}}              & $\bm{\xi_B}$ & $\bm{K_\infty}$ & $\bm{K_\mathrm{FSE}}$\\
        & \textbf{(SNU)} & \textbf{(MBit/s)} & \textbf{(MBit/s)}\\
        \hline
        \gls{voa} $0\;\si{\kilo\meter}$ & $0.0095$ & $22.4$ & $17.7$ \\
        \gls{voa} $5\;\si{\kilo\meter}$ & $0.0091$ & $11.9$ & $5.82$ \\
        \gls{voa} $10\;\si{\kilo\meter}$& $0.0076$ & $6.35$ & $2.55$ \\
        \gls{voa} $25\;\si{\kilo\meter}$& $0.0062$ & $1.43$ & $0$ \\
        Fiber $25\;\si{\kilo\meter}$    & $0.0072$ & $1.17$ & $0$ \\
    \end{tabular}}
    \caption{Average results of the experiments. $K_\infty$ is the asymptotic secret key rate and $K_\mathrm{FSE}$ is the secret key rate considering finite-size effects.}
    \label{tab:results}
\end{table}

The average results are summarized in table~\ref{tab:results}. The secret key rates are given for both the asymptotic and finite-size case. For the experiments with the fiber spool and the \gls{voa} at $25\;\si{\kilo\meter}$, the finite-size key rate is zero using the worst-case estimators. However, it is also possible to measure the finite-size key rate for each frame using the framework from~\cite{FiniteSizeAnaLeverr2010}, similarly to how the system would operate in an operational setting, but once again it is not possible to extract a secret key rate for the number of symbols used here. We obtain a positive average when considering at least $10^7$ symbols (average finite-size key rate in order of hundreds of $\si{\kilo\bit\per\second}$), but it would be challenging to reach this block size over the same number of frames, since it would roughly multiply the \gls{dsp}, and hence the experiment time, by 10. Increasing the block size will require to perform optimisation on the overall software in order to reduce the computation time (and memory usage).

\section{Discussion\label{sec:discussion}}

Using as a starting point our obtained results, there are several ways these can be extended and improved. We discuss some directions below.

\paragraph{Finite-size effects.} Unfortunately the experiments at $25\;\si{\kilo\meter}$ do not not give any key rate in the finite-size case with the worst-case estimator method. As was briefly mentioned above, another method is to follow the analysis in~\cite{FiniteSizeAnaLeverr2010} for each frame. For $N=10^6$, which was the number of symbols in our frame, we unfortunately also got a key rate of zero for each frame. However, increasing $N$ for the analysis, we start to get some frames with a positive finite-size key rate. {This is done by simply changing $N$ for the correction terms assuming the same estimated values for $\xi$ and $T$, and hence represents what we could get if we were able to obtain the same numbers with larger frames.} For the experiment with 25~km, 40 frames out of 100 at $N=10^7$ and then all the frames at $N=10^8, 10^9, 10^{10}$. For the experiment with the fiber, there were respectively 14, 77, 93 and 94 frames with a positive finite-size key rate at $N=10^7, 10^8, 10^9, 10^{10}$. We now give the average key rate over all the frames (and in parentheses the average over the positive frames) for the \gls{voa} experiment at 25~km: $96\;\si{\kilo\bit\per\second}$ ($240\;\si{\kilo\bit\per\second}$), $817\;\si{\kilo\bit\per\second}$, $1.1\;\si{\mega\bit\per\second}$, $1.2\;\si{\mega\bit\per\second}$ for $N=10^7, 10^8, 10^9, 10^{10}$ and for the experiment with the fiber: $39\;\si{\kilo\bit\per\second}$ ($280\;\si{\kilo\bit\per\second}$), $475\;\si{\kilo\bit\per\second}$ ($617\;\si{\kilo\bit\per\second}$), $718\;\si{\kilo\bit\per\second}$ ($772\;\si{\kilo\bit\per\second})$, $803\;\si{\kilo\bit\per\second}$ ($854\;\si{\kilo\bit\per\second}$) with the same $N$. This makes us confident that one way to get better results would be to increase the number of points that are used for the estimation of $T$ and $\xi$. We indeed checked that the variance of those two variables has a dependence over the number of samples used to estimate them. This would reduce the confidence interval and give better worst-case estimators and subsequently better key rates.

\paragraph{Increasing the distance.} One of the current limitations of the QOSST software is the achievable distance. In practice the \gls{dsp} can work at least until $10\;\si{\deci\bel}$ but at that value the estimation of the distance is imprecise and the excess noise too high to get a secret key. However, the distance could be improved either by improving the \gls{dsp} or by changing the generation of the pilots. Indeed, we believe that one of the limitations comes from the finite amplitude difference that can be applied between the quantum symbols and the pilots. A solution could be to generate the pilots and the quantum symbols on different paths, or to send the pilots on the other polarisation, which would allow for more powerful pilots and less crosstalk, as suggested in \cite{Pi2023}.

\paragraph{Stability for polarization.} Another important issue is the manual polarisation controller, since it allows only for a few hours of stability of the setup before another calibration becomes necessary. Two solutions may be used to handle this issue: the first is to use an automatic polarisation controller with an error signal coming from the detection to actively correct for polarisation drifts in the fiber, and the second one is to perform a polarisation-diverse detection by adding a \gls{pbs} on the signal path and detecting both polarisations by using two balanced detectors. The first one has the advantage of requiring an optical scheme very similar to the one already implemented, and of using the same \gls{dsp}. However, the algorithm to adjust the polarisation controller should be added. The second solution requires more hardware but can be easily adapted for using a double polarisation in the transmitter setup.

Since the experiments presented in this paper were done, automatic polarisation compensation with a motorised polarisation controller was implemented and tested, and is released with QOSST. {This automatic compensation algorithm showed stable recovery over tens of hours with a fiber spool.}

\paragraph{Increasing the symbol rate.} The choice of the optimised parameters reported in this work was performed using a slightly older system with a \gls{dac} that was limited to $500\; \si{\mega\sample\per\second}$ (Keysight M330A) and with a bandwidth of $200\;\si{\mega\hertz}$. This justifies the relatively low symbol rate used in the present work. The increase of the symbol rate (and hence the occupation bandwidth), allowed by the new \gls{dac} will be the subject of future investigation. We believe that there are no particular limitations coming from our software other than longer time and larger memory occupation for the \gls{dsp}.

\paragraph{Duration of the Digital Signal Processing.} Currently, the \gls{dsp} (and the parameter estimation) is run for each frame before requesting the next frame, which means that two frames are temporally separated by the \gls{dsp} time. We conducted an experiment on 20 frames to evaluate the \gls{dsp} time and we measured two durations: the first is the \gls{dsp} itself, which outputs the symbols for each subframe with a global phase difference, and the second is recovery of the global phase, which also requires Alice sending her data to Bob. The experiment is conducted with the same parameters as the experiment for $0\;\si{\kilo\meter}$ and requests $50\%$ of the symbols to Alice ($500 000$ symbols per frame). The average time for the \gls{dsp} is $70\;\si{\second\per\frame}$ and the average time for the symbol exchange and global phase correction is $164\;\si{\second\per\frame}$ for a total time between frames of $234\;\si{\second\per\frame}$. The second time can be reduced by exchanging less data for the parameter estimation. For example, taking $10\%$ of the data (\textit{i.e.} $100000$ symbols for the parameter estimation), the \gls{dsp} time is similar ($72\;\si{\second\per\frame}$ on average) and the data exchange and global phase recovery has an average time of $101\;\si{\second\per\frame}$, yielding a total of $173\;\si{\second}$ between the frames. This has the advantage of reducing the time and increasing the key rate, but the estimators for the finite-size analysis are worse. Both times could be reduced by optimizing the code, or by using a different programming language.

\section{Conclusion\label{sec:conclusion}}

We have presented a highly modular open source platform for experimental continuous-variable quantum key distribution that is able to exchange a secret key rate at metropolitan distances using the OSSB-GCMS-GG02 protocol with frequency multiplexed pilots and RF-heterodyne coherent detection. We obtain key rates of the order of the $\si{\mega\bit\per\second}$ for distances up to $25\;\si{\kilo\meter}$ at the asymptotic limit. Our platform can be easily extended to other \gls{cvqkd} protocols and hardware configurations, including with double polarisation and phase-diverse heterodyne detection. Our platform can be improved and extended by integrating techniques accessible in the near term.

This software is released to the community in the hope that it will help to stimulate further experimental research in \gls{cvqkd} by providing the initial means to set up a working \gls{cvqkd} setup. We also hope that the community will help this software suite grow, for instance by adding other \gls{cvqkd} configurations, increasing the execution speed, complementing the package for computing secret key rates under different assumptions and implementing the post processing operations.

\section*{Author contributions}
VMA implemented the global phase recovery algorithm. VMA and MS contributed to increasing the achievable distance and to tests of the software. YP implemented the rest of the software. MS, BG, LTV, PG, AR and ED participated in crucial discussions that directly contributed to the platform design and its improvements.

\section*{Acknowledgments}

Figure \ref{fig:cvqkd_schema} was created by partly using the gwoptics component library distributed under CC-BY-NC license, attribution to Alexander Franzen \cite{gwoptics}. The authors acknowledge financial support from the European Unions’s Horizon Europe research and innovation programme under the project QSNP (Grant No. 101114043) and the PEPR integrated project QCommTestbed (ANR-22-PETQ-0011), which is part of Plan France 2030.

\bibliographystyle{quantum}
\bibliography{bibliography, others, software}

\onecolumn
\appendix

\section{Calibration of Alice\label{appendix:calibration-alice}}

The calibration of the optical setup of Alice consists mainly in the calibration of the conversion factor $r_{\mathrm{conv}}$. This is done by directly connecting the input of Alice's \gls{voa} to the laser, and adding a powermeter at the output of Alice. While the power is kept constant, the attenuation of the \gls{voa} is changed and, for each attenuation, the power on the monitoring photodiode and the one on the powermeter at Alice's output are recorded and $r_{\mathrm{conv}}$ is measured by a linear fit of the output power to the monitoring power.
This method has several advantages: first $r_{\mathrm{conv}}$ doesn't depend on the \gls{voa} (this is due to the fact that the monitoring photodiode is after the \gls{voa}) and $r_{\mathrm{conv}}$ is estimated without changing the optical connection (the only change in the optical connections happens before the region of $r_{\mathrm{conv}}$). In our setup the measured value of $r_{\mathrm{conv}}$ was around $4.4\cdot 10^{-3}$, which is compatible with $r_{\mathrm{conv}}^{\mathrm{ideal}} = \frac{0.05}{0.95}\cdot 0.1 \simeq 5.3\cdot 10^{-3}$.

A characterization of Alice's \gls{voa} is necessary to know the relation between the applied voltage and the attenuation. In the experiment, a voltage of $2.9\;\si{\volt}$ is applied, corresponding to an attenuation of $3.78\;\si{\deci\bel}$.
Both characterizations can be done using a script in the qosst-alice package.

\section{Calibration of Bob\label{appendix:calibration-bob}}

The calibration of Bob is mainly composed of the measurement of the detector efficiency $\eta$, which takes into account the global efficiency of the receiver on the quantum signal path (from the input of the quantum signal to the photodiodes).
For a single photodiode, the efficiency is defined as the ratio between the number of emitted electrons and the number of received photons, which reaches $\eta_{PD}=1$ for the ideal photodiode that emits one electron for each incoming photon. One can relate the efficiency of a photodiode to its responsivity $\mathcal{R} = I/P$ (\textit{i.e.} the ratio between the photocurrent and the optical power) as
\begin{equation}
    \eta_{PD} = \frac{n_e}{n_p} = \frac{\frac{I}{e}}{\frac{P}{\frac{hc}{\lambda}}} = \frac{hc}{\lambda e} \mathcal{R} \stackrel{\lambda = 1550\;\si{\nano\meter}}{\simeq} \frac{\mathcal{R}}{1.25\;\si{\ampere\per\watt}}.
\end{equation}

For a balanced detector, the responsivity is usually defined as the sum of the two photocurrents divided by the input power at the beginning of the interferometer. For the whole detector, we can compute the responsivity as the sum of all photocurrents divided by the input optical power, and then divide this value by $1.25\;\si{\ampere\per\watt}$ to get the efficiency.
In our setup, the Thorlabs balanced detector has two voltage monitoring outputs $V_+, V_-$, such that
\begin{equation}
    \begin{split}
        V_{+} &= G_m\cdot I_+, \\
        V_{-} &= G_m\cdot I_-,
    \end{split}
\end{equation}
where $I_+$ and $I_-$ are the photocurrents, $G_m = 10\;\si{\kilo\volt\per\ampere}$ is the monitoring gain and $V_+$ and $V_-$ are the voltages of the monitoring outputs. Hence
\begin{equation}
    \eta = \frac{hc}{\lambda e} \cdot \frac{V_+ + V_-}{G_m \cdot P_{sig}},
    \label{eqn:eta}
\end{equation}
where $P_{sig}$ is the optical power at the signal input of Bob. In practice, this calibration is done by putting a laser at a power of a few $\si{\milli\watt}$, followed by a \gls{voa}, and a 50:50 beam splitter with one output on a powermeter and the other on the signal input of Bob. The attenuation of the \gls{voa} is changed and for each attenuation the value of $V_+$, $V_-$ and $P_{sig}$ are recorded. The detector efficiency $\eta$ is measured by linear fit using the formula in Eq.~(\ref{eqn:eta}).

The value of $\eta$ changes between the setup with the emulated channel using the \gls{voa} and the one with the fiber spool, since in the latter a manual polarisation controller is added to Bob's setup. For both cases, $\eta$ is estimated using an automated script available in the qosst-bob package, giving $\eta_{voa} = 55.4\%$ and $\eta_{fiber}=41.3\%$. We can compare those values with the expected efficiency of the receiver. The responsivity of the photodiodes in the balanced detector is typically around $0.9\;\si{\ampere\per\watt}$ at $1550\;\si{\nano\meter}$ (verified by measurement of the balanced detector only), giving an efficiency of $\eta_{PD} = 72\%$. The excess losses are typically 7\% (beam splitter), 15\% (switch), 10\% (polarisation maintaining mating sleeves, 2 in total), giving $\eta_{typ} = 46\%$.
Monitoring the photocurrents is crucial in order to get a good estimation of the detector efficiency $\eta$, which can lead to security issues if not properly estimated, in particular in the trusted detector model.

The receiver is also characterized by another important parameter, which is the electronic noise $V_{el}$ (or clearance). The precise value of $V_{el}$ needs to be calibrated for each frame for two reasons: first, the shot noise is calibrated at each step, and as $V_{el}$ is normalised relative to the shot noise, the value changes with the shot noise, and, second, $V_{el}$ is dependent on the value of $f_{beat}$ (and the other frequency parameters). Indeed, the same \gls{dsp} is applied to the electronic noise samples (and the electronic and shot noise samples) before measuring the noises, and the exact behaviour of the \gls{dsp} depends on the frequency parameters.

\section{Shot noise units and normalisation\label{appendix:snu}}

In natural units, the following relations hold in terms of the quadrature operators:
\begin{equation}
    \begin{split}
        [\hat{q}, \hat{p}] &= i\hbar,\\
        \Delta \hat{q}\Delta \hat{p} &\geq \frac{\hbar}{2}.
    \end{split}
\end{equation}
The \glsxtrfull{snu} are defined by choosing $\hbar=2$ so that
\begin{equation}
    \begin{split}
        [\hat{q}, \hat{p}] &= 2i,\\
        \Delta \hat{q}\Delta \hat{p} &\geq 1.
    \end{split}
\end{equation}
Using this convention, the shot noise is equal to 1. In practice, we normalize the received quantum symbols by the variance of the shot noise symbols. The electronic and electronic and shot noise symbols are obtained by applying the same \gls{dsp} on the quantum, the electronic noise and the electronic and shot noise, giving the quantum symbols, the electronic symbols and the electronic and shot noise symbols. The variance of the shot noise is obtained and used for normalization to ensure that the variance of the shot noise is 1.

\section{RF-Heterodyne\label{appendix:rf-heterodyne}}

RF-heterodyne is a detection scheme that only uses one balanced detector to measure both quadratures. The basic idea is that the data is encoded on a single sideband, by displacing it with a frequency $f_{\mathrm{shift}} > \frac{B}{2} = \frac{(1+\beta)\cdot R_s}{2}$. In this way, the data can be recovered by detecting the signal after a physical homodyne detection and recovering the two encoded quadratures by demodulating the signal by multiplying by the complex exponential at frequency $-f_{\mathrm{shift}}$. In practice, the demodulation frequency is not exactly $-f_{\mathrm{shift}}$ but $-f_{\mathrm{shift}} - f_{\mathrm{beat}}$, where $f_{\mathrm{beat}}$ is the frequency difference between the two lasers.

This method has some limitations: one is limited to modulating only one sideband, and needs to ensure that no information is leaking to the other, unmonitored sideband. However, the setup is much simpler, using only one balanced detector, which also simplifies the analysis on the shot and electronic noise. It adds however some complexity in the \gls{dsp} with the demodulation.

\section{Digital Signal Processing\label{appendix:dsp}}

\subsection{Zadoff-Chu sequence}

Here, we extend the explanation of the \gls{dsp} algorithm, and give a scheme with some examples.
The Zadoff-Chu sequence is defined by the following formula:
\begin{equation}
    ZC(n)=\exp\left(-j\frac{\pi R_{ZC}n(n+c_\text{f}+2q)}{L_\text{ZC}}\right),
\end{equation}
with $0\leq n < N_{ZC}$, $R_{ZC}$ the root of the sequence, $L_{ZC}$ the length of the sequence, $q$ the cyclic shift and $c_\text{f} = N_{ZC} \;\text{mod}\;2$. The Zadoff-Chu sequence requires that $0\leq R_{ZC} \leq L_{ZC}$ and that $R_{ZC}$ and $L_{ZC}$ are coprimes.
In practice, we choose $L_{ZC}$ and $N_{ZC}$ to be prime numbers and $q=0$, which simplifies the generation formula of the Zadoff-Chu sequence,
\begin{equation}
    ZC(n)=\exp\left(-j\frac{\pi R_{ZC}n(n+1)}{L_\text{ZC}}\right).
\end{equation}
This sequence is called a \glsxtrfull{cazac} sequence and its good periodic autocorrelation properties make it extremely fit for time synchronisation.\\

\subsection{Raised Cosine and Root-Raised Cosine}

\glsxtrfull{rc} filters are a family of filters that minimize \glsxtrfull{isi}, defined by the following frequency domain description
\begin{equation}
    H_{rc}(f)=\begin{cases}
        1,
         & |f| \leq \frac{1 - \beta}{2}\cdot R_S                                \\
        \frac{1}{2}\left[1 + \cos\left(\frac{\pi}{\beta\cdot R_S}\left[|f| - \frac{1 - \beta}{2}\cdot R_S\right]\right)\right],
         & \frac{1 - \beta}{2}\cdot R_S < |f| \leq \frac{1 + \beta}{2}\cdot R_S \\
        0,
         & \text{otherwise},
    \end{cases}
\end{equation}
where $\beta$ is the roll-off of the \gls{rc} filter and $R_S$ is the symbol rate.
The bandwidth of this filter is $2\cdot\frac{1+\beta}{2}\cdot R_S = (1+\beta)\cdot R_S$.
Using the time description of the filter, it is possible to see that a perfect sampling will induce to recover one symbol with no interference from any other symbol.
In practice, we want to apply a filter at the transmitter side and a matched filter at the receiver side to optimise the \gls{snr}, as the matched filter is the filter that maximizes the \gls{snr} in the presence of white noise. For this we use the \glsxtrlong{rrc} filter defined as
\begin{equation}
    H_{rc}(f) = H_{rrc}(f) \cdot H_{rrc}(f).
\end{equation}

\subsection{Scheme of the Digital Signal Processing}

\begin{figure*}
    \centering
    \includegraphics[scale=0.6]{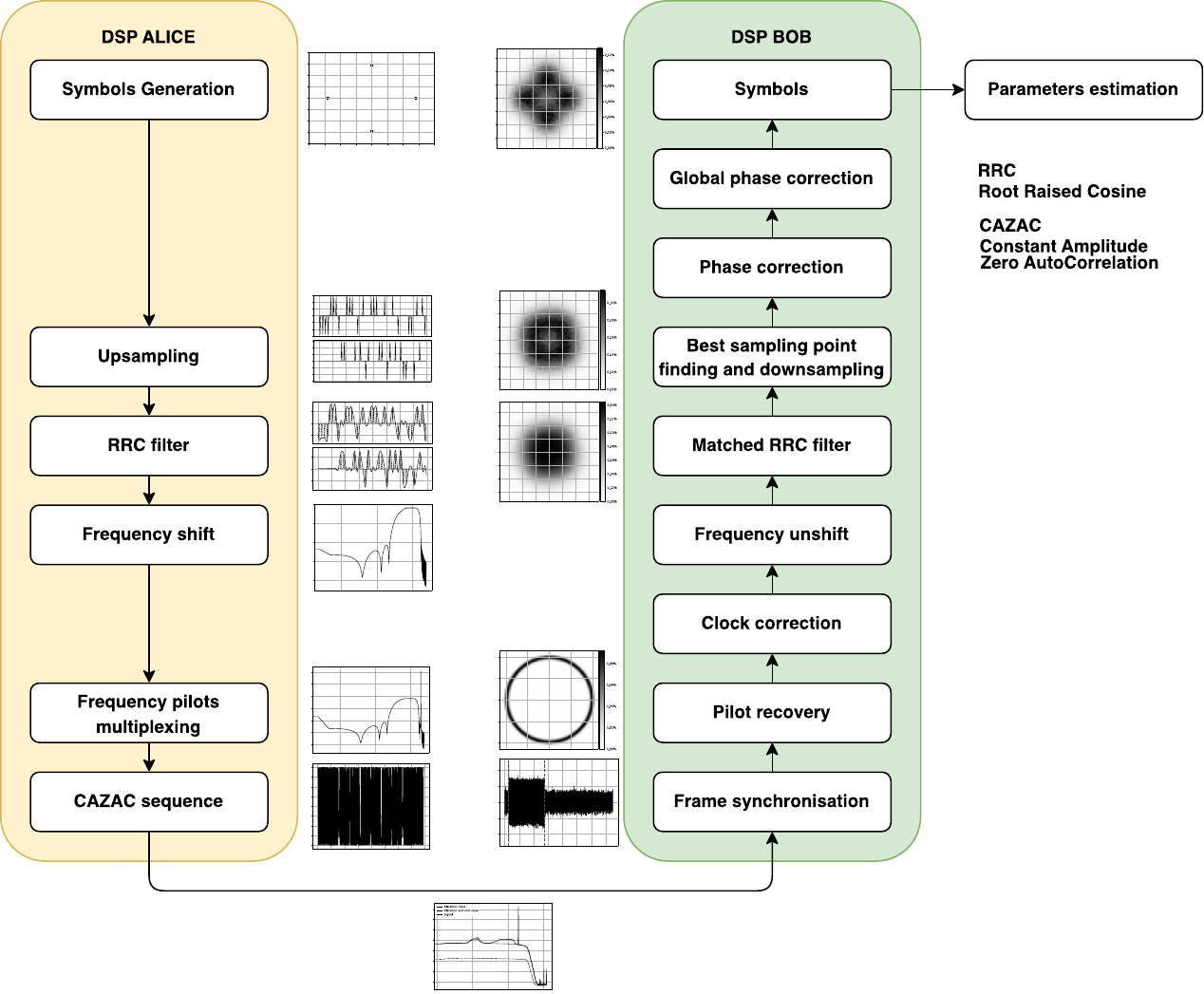}
    \caption{Scheme of the Digital Signal Processing in QOSST.}
    \label{fig:dsp}
\end{figure*}

In Fig.~\ref{fig:dsp} we provide a scheme of the signal processing applied in the QOSST software. The \gls{qpsk} modulation is used only for display purposes.

\end{document}